\documentclass[twocolumn,11pt]{ndes09}

\usepackage{times}

\usepackage{amsmath}
\usepackage{amssymb}
\usepackage{epsfig}
\usepackage{float}
\usepackage{cite}

\newcommand{\etal}{{\it et al.}}

\begin{document}


\title{Regular spiking in asymmetrically delay-coupled FitzHugh-Nagumo
  systems}
\author{{\it A. Panchuk$^{a, 1}$, M. Dahlem$^{b, 2}$,
    E. Sch\"{o}ll$^{b, 3}$}\\[2ex]  
$^{a}$Institute of Mathematics, \\ 
National Academy of Sciences of Ukraine \\ 
Kyiv, Ukraine \\ 
$^{b}$Institut f\"{u}r Theoretische Physik, \\
Technische Universit\"{a}t Berlin, \\
Berlin, Germany \\
$^{1}$e-mail: {\tt nastyap@imath.kiev.ua} \\
$^{2}$e-mail: {\tt dahlem@physik.tu-berlin.de} \\
$^{3}$e-mail: {\tt schoell@physik.tu-berlin.de}}

\maketitle


\thispagestyle{empty}
\pagestyle{empty}



\begin{abstract}
We study two delay-coupled FitzHugh-Nagumo systems, introducing
a mismatch between the delay times, as the simplest
representation of interacting neurons. We demonstrate that the
presence of 
delays can cause periodic oscillations which coexist
with a stable fixed point. Periodic solutions observed are of two
types, which we refer to as a ``long'' and a ``short'' cycle,
respectively.
\end{abstract}


\section{Introduction}

Being an inherent feature of a human, laziness was always that force
which engendered invention of new devices, supposed to work instead
of people. And in our epoch of vast technological progress, there are
thousands of 
useful gadgets already existing. Though recently the
science is advancing with 
seven-league strides, there are still a great number of phenomena
which are waiting 
for 
a better insight. 
One has to admit that none of the existing complex machines and
powerful computers can substitute 
a single human brain. Which means that we still do not draw close
enough to clearing up a mystery of how this accumulation of grey
matter really works.  

Since the end of the last century, study of neural networks picks up
speed.
In order to describe its intricate behavior, the brain is often
represented as an ensemble of coupled nonlinear dynamical elements,
capable of producing spikes and exchanging information between each
other \cite{hak06, wil99, ger02}. Such neural populations are usually
spatially localized and 
contain both excitatory and inhibitory neurons \cite{wil72}.

Some researchers, starting from the simplest case of two
interconnected neurons, show how more complicated dynamics emerges in
larger sets \cite{des94}. The others explore  extremely complex network of
subnetworks, focusing on the hierarchically clustered organization of
interacting excitable elements \cite{zho06}. 

Most studies base on the present oscillatory behavior of
individual system elements, which then produces observable patterns
due to collective synchronization \cite{ros04, pop05a, gas07}. Thus, for
modeling a single neuron, phase oscillators are often
used. For
instance, to characterize mutual dynamics of cells in certain
brain areas, responsible for giving the onset to Parkinson's disease
or epilepsy, a well-known Kuramoto model is considered \cite{pop05b, ash06,
mai07, ash08}.

Here, we rely on the works by FitzHugh
\cite{fit61} and Nagumo \etal \cite{nag62} who have shown that for
describing the main  
characteristics of a neuron dynamics, it is sufficient to
consider a 2-dimensional system. The latter is also widely used
nowadays as one of the simplest models for examining brain dynamics
and has been essentially studied in many papers (see, for instance,
\cite{bur03, lin04, tor03, kit05}
and references therein). 

Having an intention to move from simple to complex, we
consider below a set of equations consisting only of two identical
FitzHugh-Nagumo subsystems (see also \cite{dah08, sch09}). Their
interaction is 
described by a linear coupling term which includes delays ($\tau_1$
and $\tau_2$), accounted for the fact that the signal transmission
between neurons is not instantaneous:
\begin{equation}
  \label{eq:fhn}
  \begin{split}
    \epsilon\dot{x}_1 &= x_1 - \frac{x_1^3}{3} - y_1 + C(x_2(t - \tau_2)
    - x_1(t)) \\
    \dot{y}_1 &= x_1 + a \\
    \epsilon\dot{x}_2 &= x_2 - \frac{x_2^3}{3} - y_2 + C(x_1(t - \tau_1)
    - x_2(t)) \\
    \dot{y}_2 &= x_2 + a
  \end{split}
\end{equation}
Here $(x_1, y_1)$ and $(x_2, y_2)$ are the phase coordinates for the
first and the second subsystem respectively. The
parameter $a$ determines whether the individual
neuron is in the excitable regime or exhibits 
self-sustained periodic firing. The time scale parameter $\epsilon$ is
chosen during the numerical simulations to be $0.01$, which results in fast
activator variables $x_1$, $x_2$ and slow inhibitor variables $y_1$, $y_2$.
For further simplicity, the coupling strength $C$ is also taken
symmetric.

\section{Sketch dynamics}

\subsection{Fixed point}
As it was already mentioned, the dynamics of an isolated 2-dimensional
FitzHugh-Nagumo system is already well-investigated. Its
single fixed point $P_2 = (-a, a^3/3 - a)$ is stable for $a > 1$ and
exhibits a supercritical Hopf bifurcation when the excitability
parameter crosses unity, which implies periodic spiking for $a <
1$. Provided that $a > 1$, the system is excitable, namely, if a
sufficient external impulse is added, it emits a spike and then rests
again in the $P_2$ state.

For our numerical simulations, we take $a = 1.3$, so that the
individual subsystems are in the excitable regime. 
The coupling term of the considered form is canceled for a fixed
point orbit, thus, the 4-dimensional equilibrium 
$P_4 = (-a, -a + {a^3}/{3}, -a, -a + {a^3}/{3})$, being existent
for the uncoupled system, persists as well for the
Eq.~(\ref{eq:fhn}). Changing the coupling strength or the delays
also does not influence its stability, as it was
recently shown \cite{dah08}.





\subsection{Regular spiking}
\begin{figure}[h]
  \centering
  \epsfig{file = {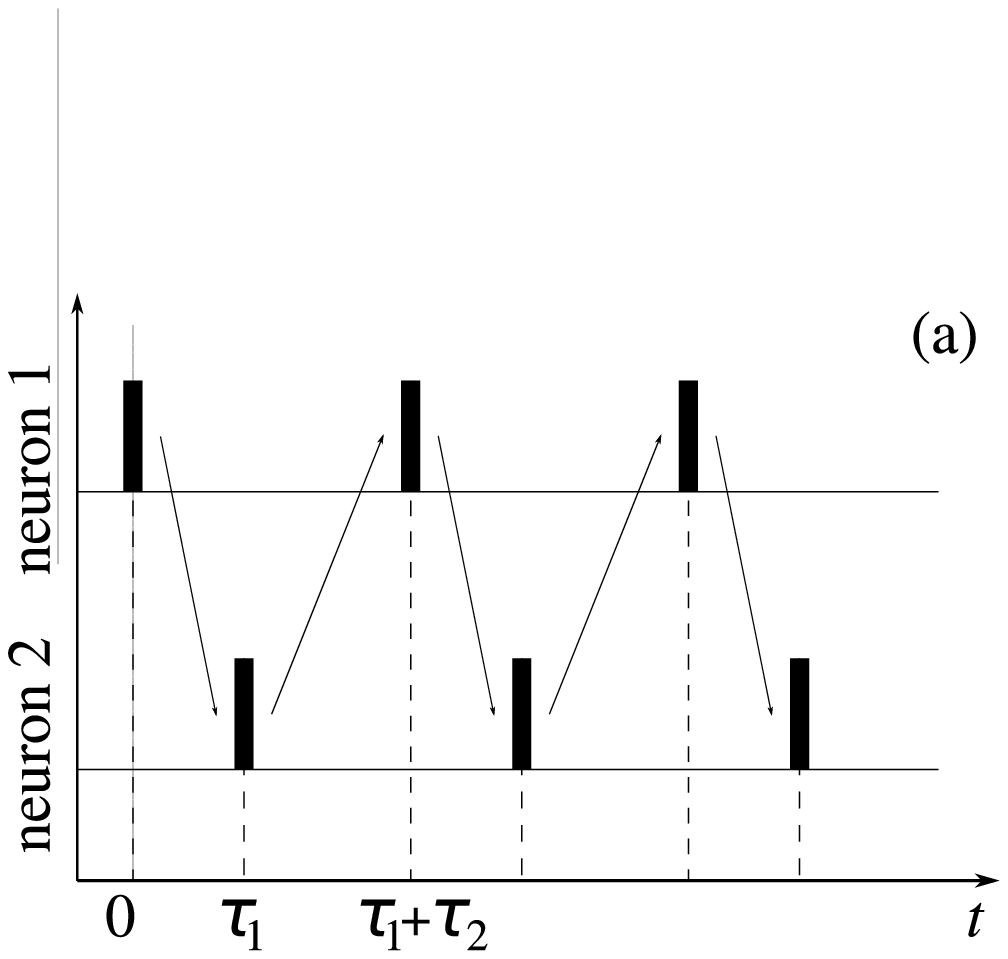}, width = 0.46\linewidth}
  \hspace{0.01\linewidth}
  \epsfig{file = {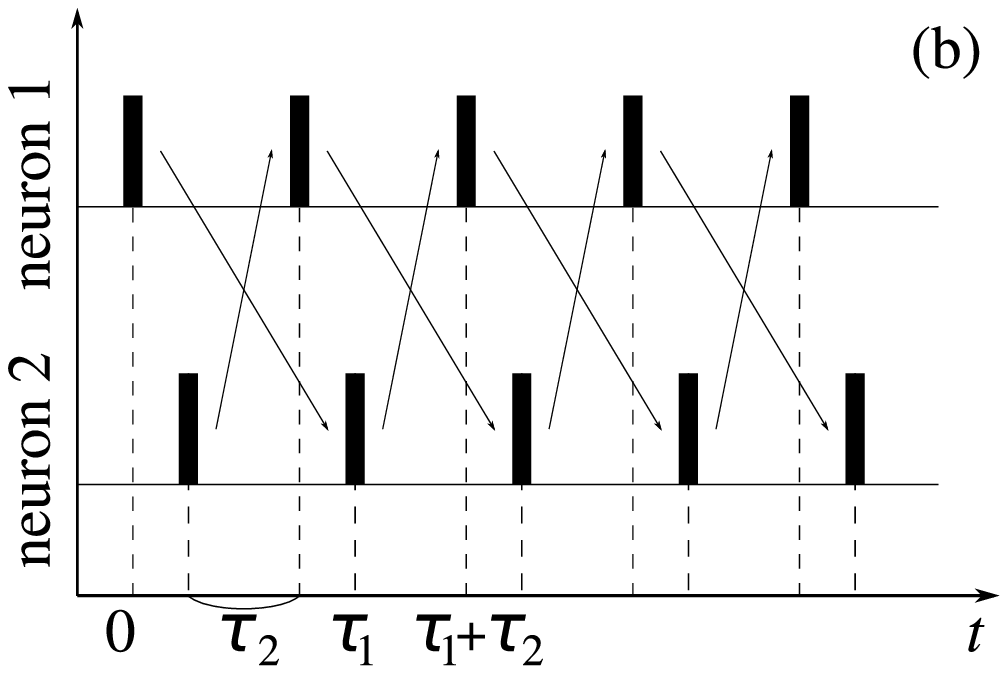}, width = 0.46\linewidth}
  \caption{Schematic representation of periodic firing in a system
    with delay. (a) ``long'' cycle; (b) ``short'' cycle}
  \label{fig:sch_lng}
\end{figure}

However, besides the stable fixed point solution, the system
(\ref{eq:fhn}) can also produce periodic oscillations. Intuitively,
this phenomenon can be explained as follows. One can perturb, for 
instance, the first neuron, so that it emits a spike. Then, with the
delay $\tau_1$ this perturbation reaches the second neuron, which
provokes it to spike as well. Again with the delay $\tau_2$ the second
neuron ``informs'' the first one that it has been stimulated, which 
causes a new run of the cycle, and the process repeats (see
schematic representation in the Fig~\ref{fig:sch_lng}(a)).

Though, in the numerical simulations, starting from various initial
conditions, 
we observed periodic solutions of {\it two} different types, which are
referred to in the following as a ``long'' and a ``short'' cycle
respectively. The former is of the period $T \gtrapprox \tau_1 +
\tau_2$, while the latter has the period $T \gtrapprox (\tau_1 +
\tau_2)/2$. 

\begin{figure}[t]
  \centering
  \epsfig{file = {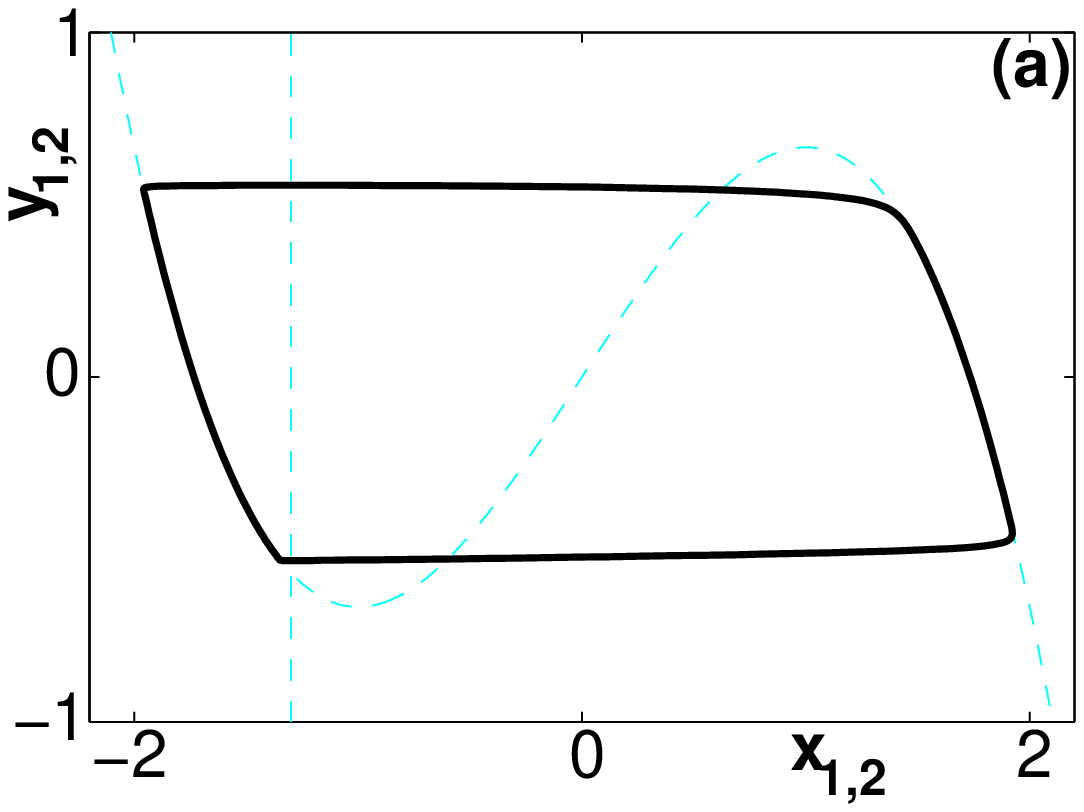}, width = 0.46\linewidth}
  \hspace{0.02\linewidth}
  \epsfig{file = {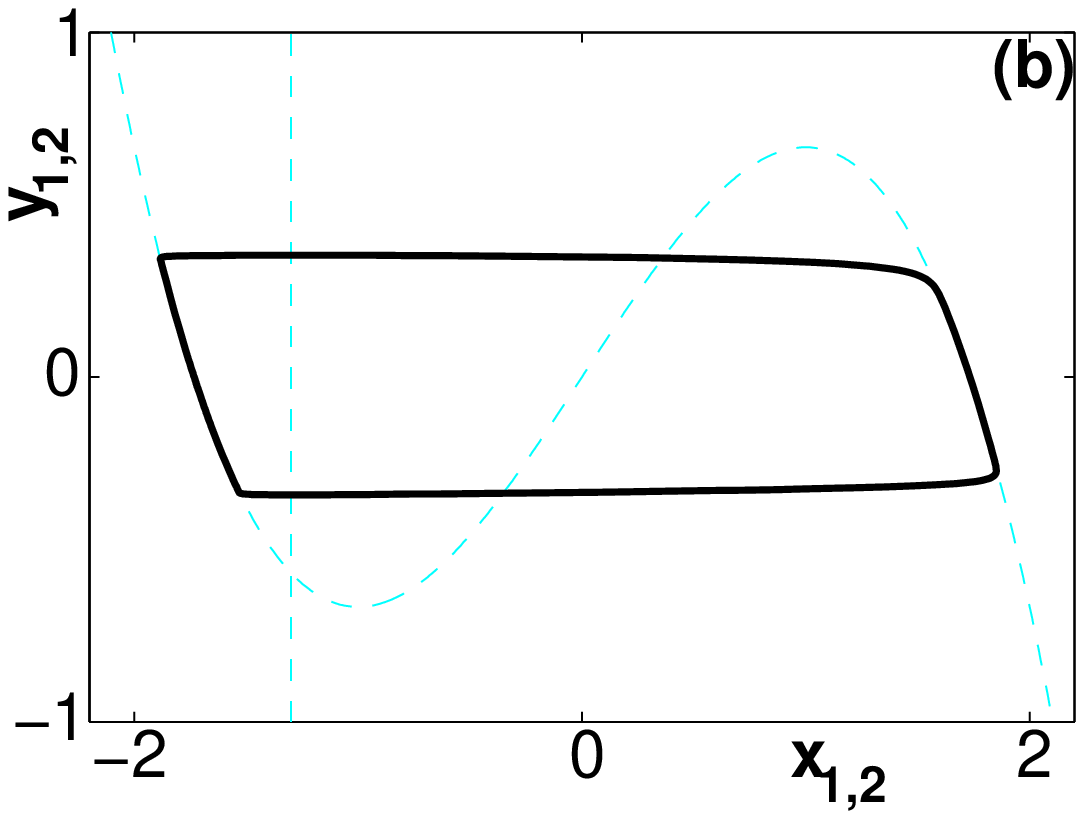}, width = 0.46\linewidth} \\
  \epsfig{file = {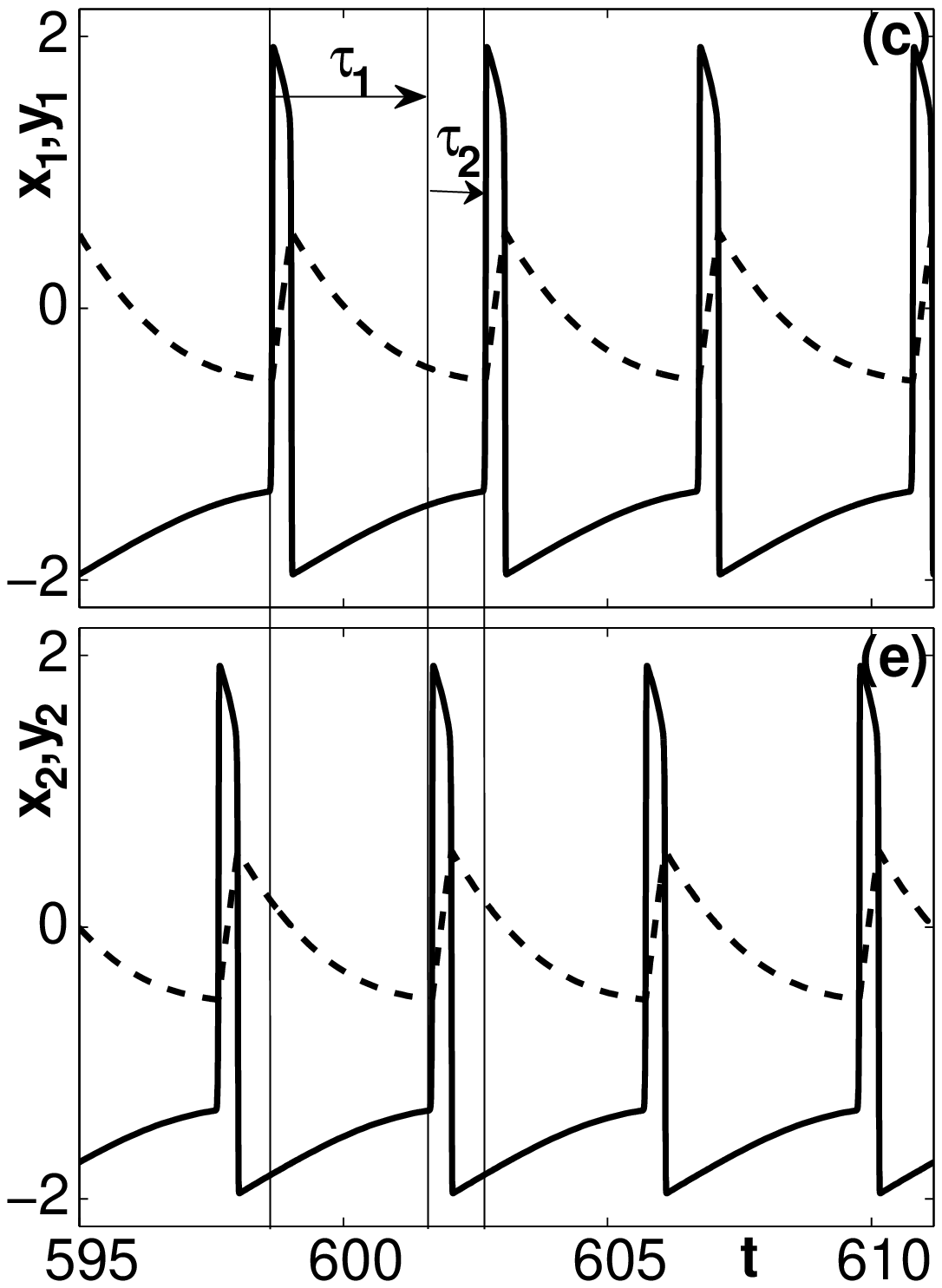}, width = 0.46\linewidth}
  \hspace{0.02\linewidth}
  \epsfig{file = {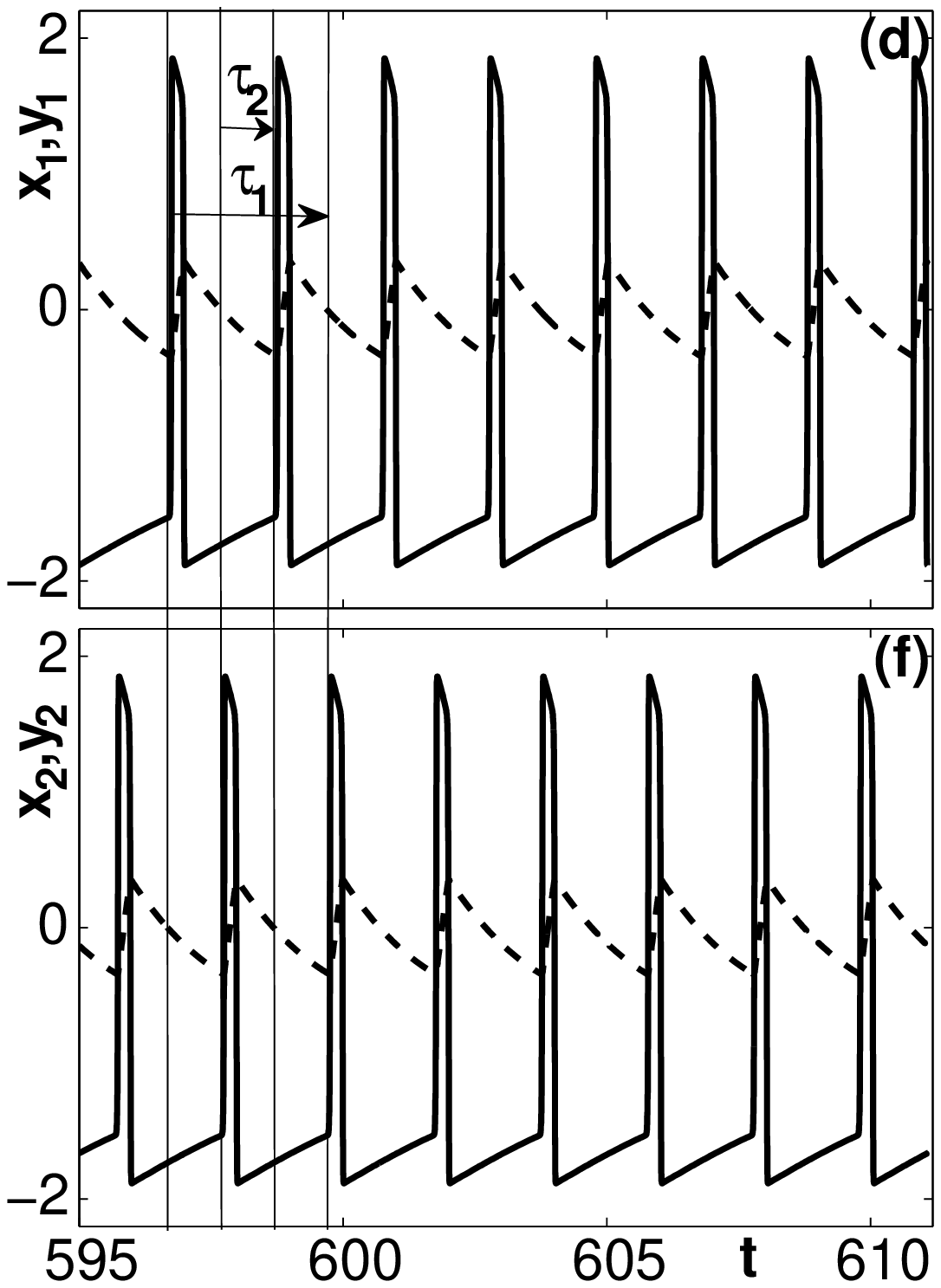}, width = 0.46\linewidth}
  \caption{Phase portraits ((a) and (b)) and time series ((c), (e) and
    (d), (f)) for long and short cycles respectively. On the time
    series plots, solid line correspond to $x_{1, 2}$ and dashed line
    correspond to $y_{1, 2}$. The parameters are $a = 1.3,
    \epsilon_{1, 2} = 0.01, C = 0.5, \tau_1 = 3, \tau_2 = 1$.}
  \label{fig:difcyc}
\end{figure}
Again, intuitively, to obtain this second solution one would add an
initial impulse not to one, but to both neurons, then roughly the
short cycle 
dynamics can be plotted as in the Fig.~\ref{fig:sch_lng}(b). 
One could remark that, in this case, the initial perturbation for the
second neuron should arrive before the delayed signal of the first
one, namely for $t \in (0, \tau_1)$. Although there are infinitely
many variations for choosing the time moment for the second impulse,
in our numerical 
simulations we were able to observe only that pattern, which is depicted
in the Fig.~\ref{fig:sch_lng}(b).

In the Fig.~\ref{fig:difcyc}, we plot the phase portraits and the
data series for these two attractors for certain fixed parameter values.

\section{Periodic solutions: deeper insight}

The next point to investigate in connection with the periodic
firing patterns obtained, is a question whether these solutions exist
for all couplings. Is their stability region large enough or such
solutions appear only for separate parameter values?
 
\subsection{``Long'' cycle}

\begin{figure}[h]
  \centering
  \epsfig{file = {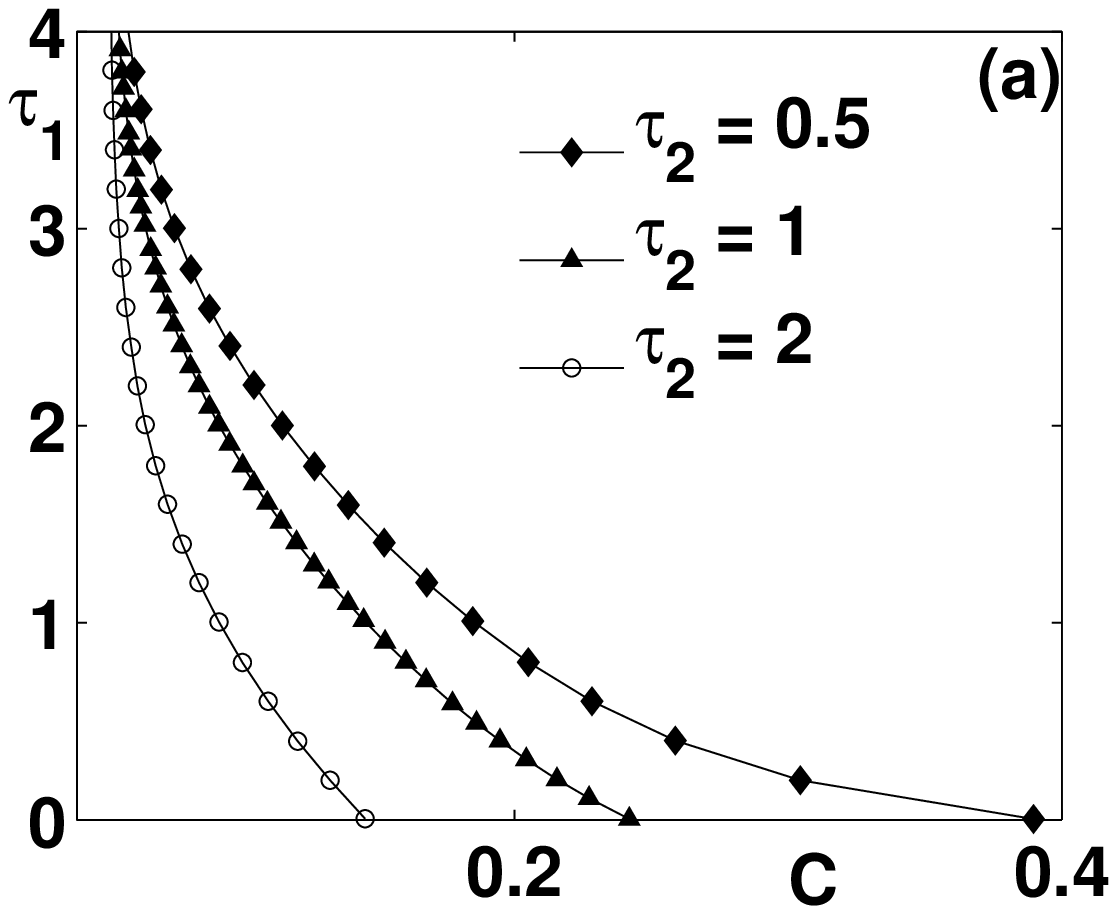}, width = 0.41\linewidth}
  \hspace{0.05\linewidth}
  \epsfig{file = {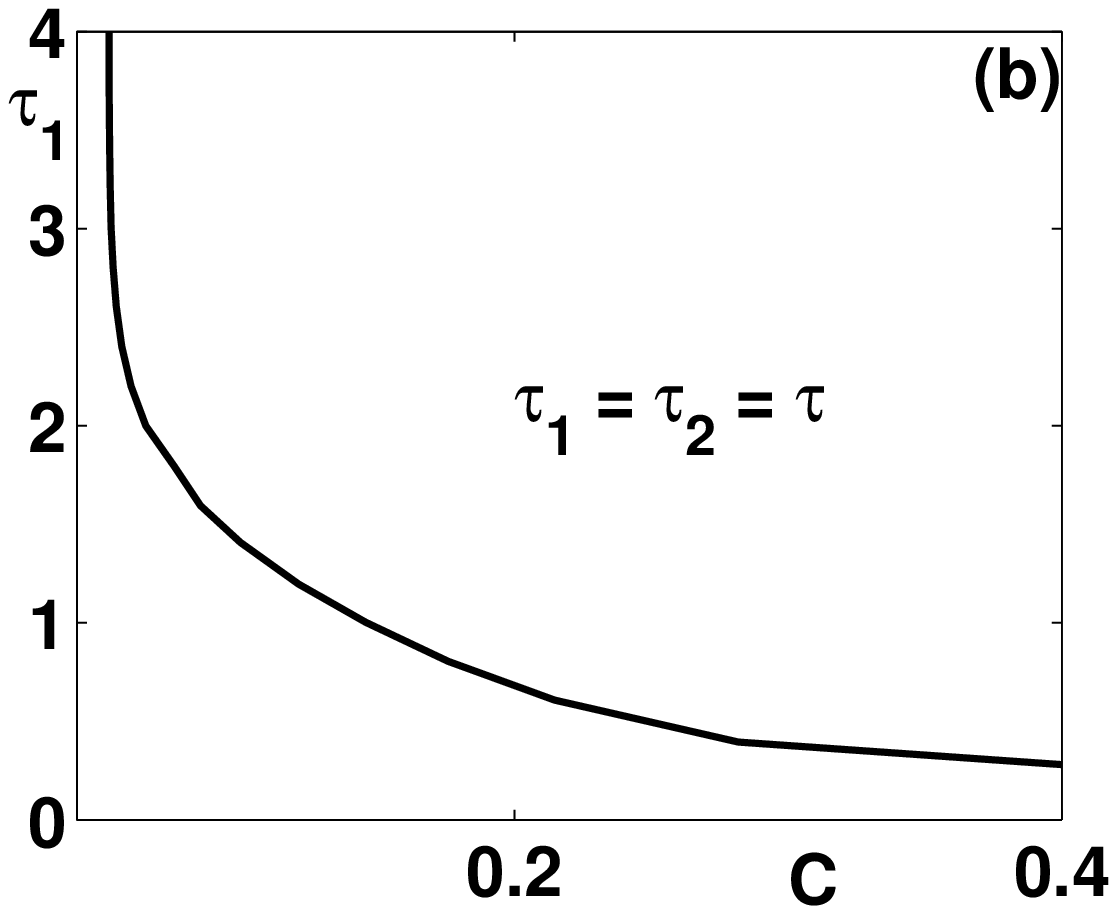}, width = 0.41\linewidth} \\
  \epsfig{file = {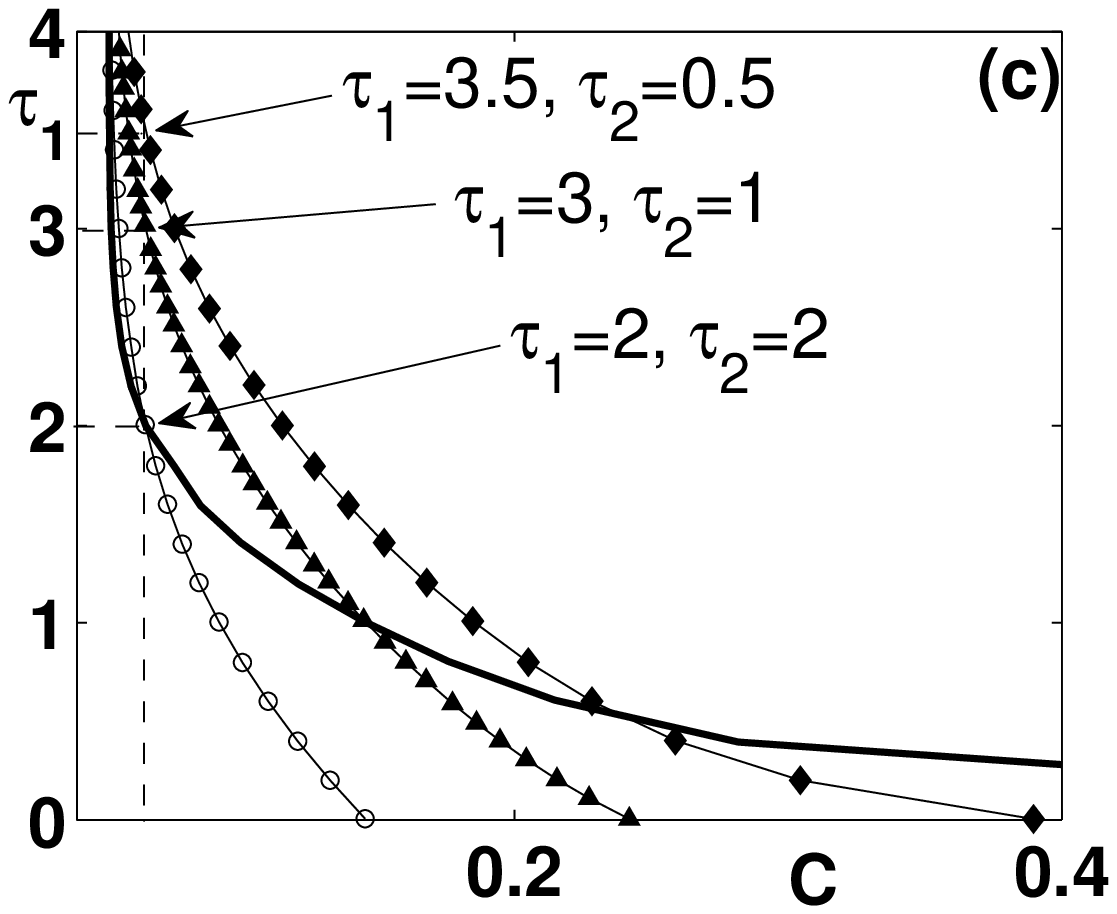}, width = 0.41\linewidth}
  \caption{Bifurcation curves for appearance of the $(\tau_1 +
    \tau_2)$-periodic solution (long cycle) 
    in $(C, \tau_1)$-plane. (a) Diagrams for different $\tau_2$
    values. (b) Diagram for the case of $\tau_1 = \tau_2 = \tau$. (c)
    Overlay of the figures (a) and (b). Vertical dashed line indicates
    a critical value of $C$.}
  \label{fig:long_bc}
\end{figure}

As it was already noticed in \cite{dah08}, such oscillations
appear through a saddle-node bifurcation of limit cycles, creating a
pair of a stable and an unstable periodic orbit.
%
In the Fig.~\ref{fig:long_bc}(a), the bifurcation curves of this
attractor type are plotted in the $(C, 
\tau_1)$-plane, for $\tau_2 = 0.5$, $\tau_2 = 1$ and $\tau_2
= 2$. It is easy to
conclude, that with increasing $\tau_2$ the bifurcation curve moves to
the left, closer to the wall value $C = 0$. 

This implies
that for some large enough coupling periodic firing still
exists even if one of the delays is close to zero.
For comparison, in the Fig.~\ref{fig:long_bc}(b), the bifurcation
curve for the case 
$\tau_1 = \tau_2$ is present. When overlaying the two graphs of (a)
and (b) (Fig.~\ref{fig:long_bc}(c)), one can notice that the critical
coupling value (indicated by a vertical dashed line) 
does not depend on the delay times difference, but only on their
sum (see Appendix). 

\begin{figure}[h]
  \centering
  \epsfig{file = {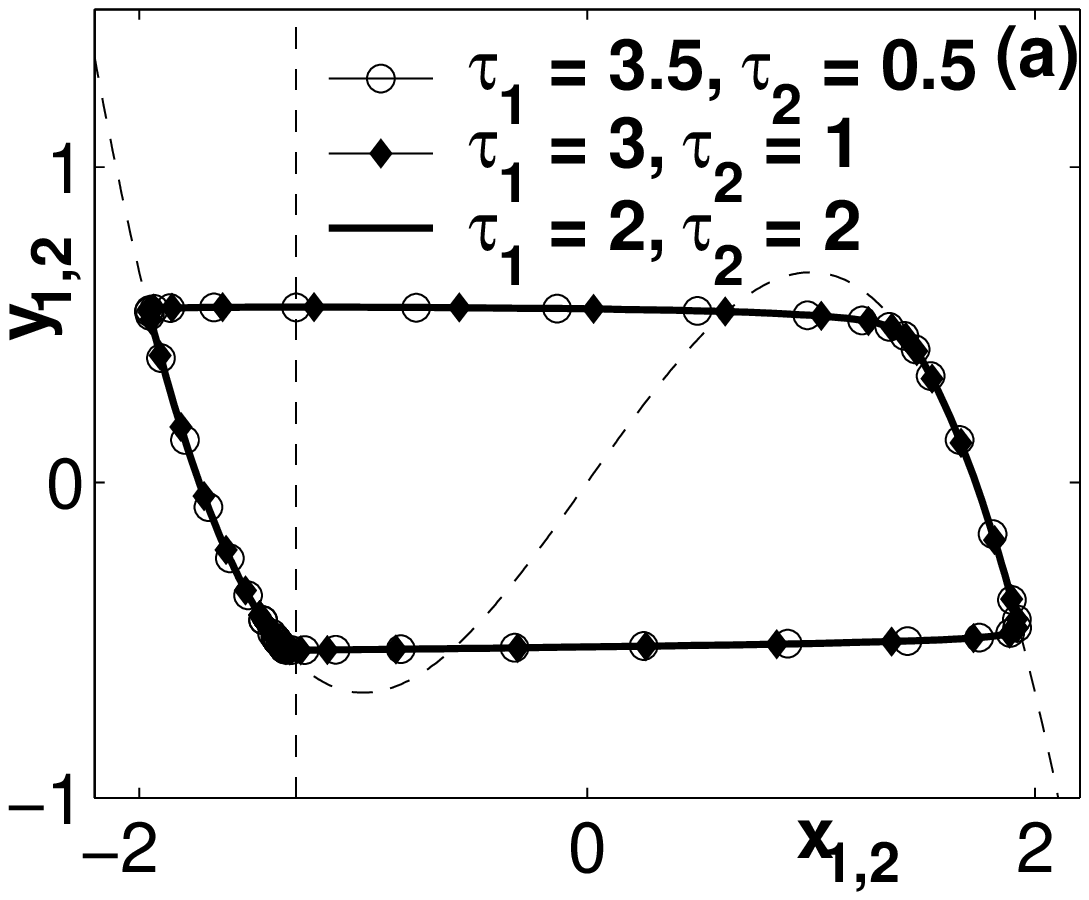}, height = 0.36\linewidth}
  \hspace{0.05\linewidth}
  \epsfig{file = {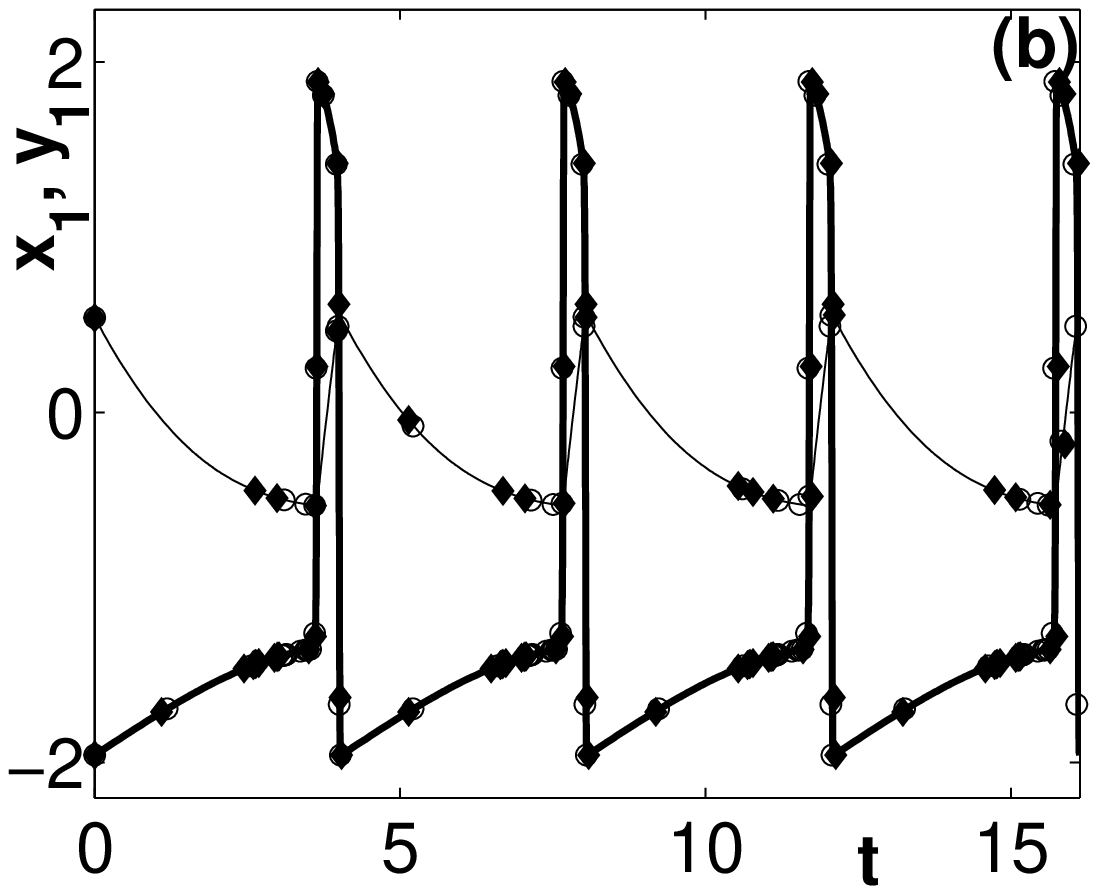}, height = 0.36\linewidth}
  \caption{Phase portrait and time series for 3 different long cycles
    with $C = 0.5$.}
  \label{fig:long_comp}
\end{figure}

In support to this last statement, we depict in the 
Fig.~\ref{fig:long_comp}(a, b) phase portraits and time series for
three different periodic solutions,  
namely
$\tau_1 = \tau_2 = 2$, $\tau_1 = 3, \tau_2 = 1$ and $\tau_1 = 3.5,
\tau_2 = 0.5$, while the sum of delays is always 4 and the coupling
strength $C = 0.5$. As it could be
clearly seen, the phase trajectories coincide perfectly as well as
the time series.

\begin{figure}[h]
  \centering
  \epsfig{file = {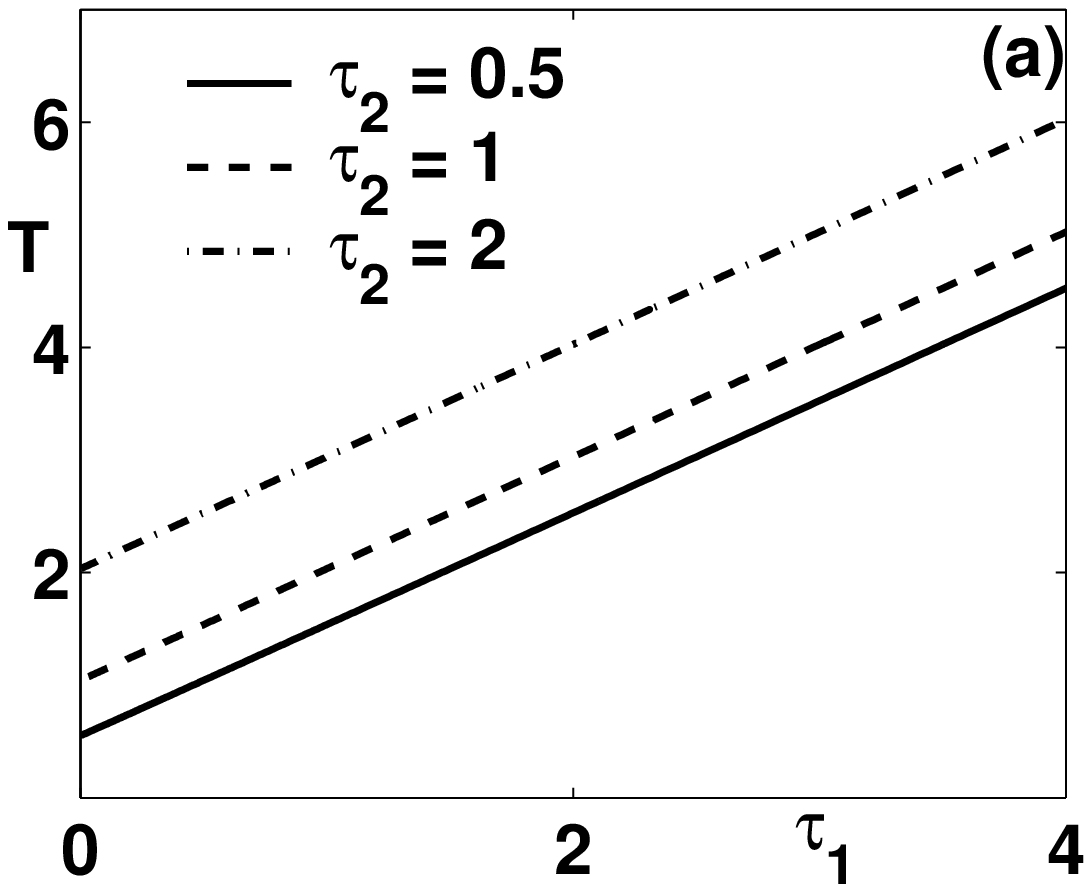}, height =
    0.335\linewidth}
  \epsfig{file = {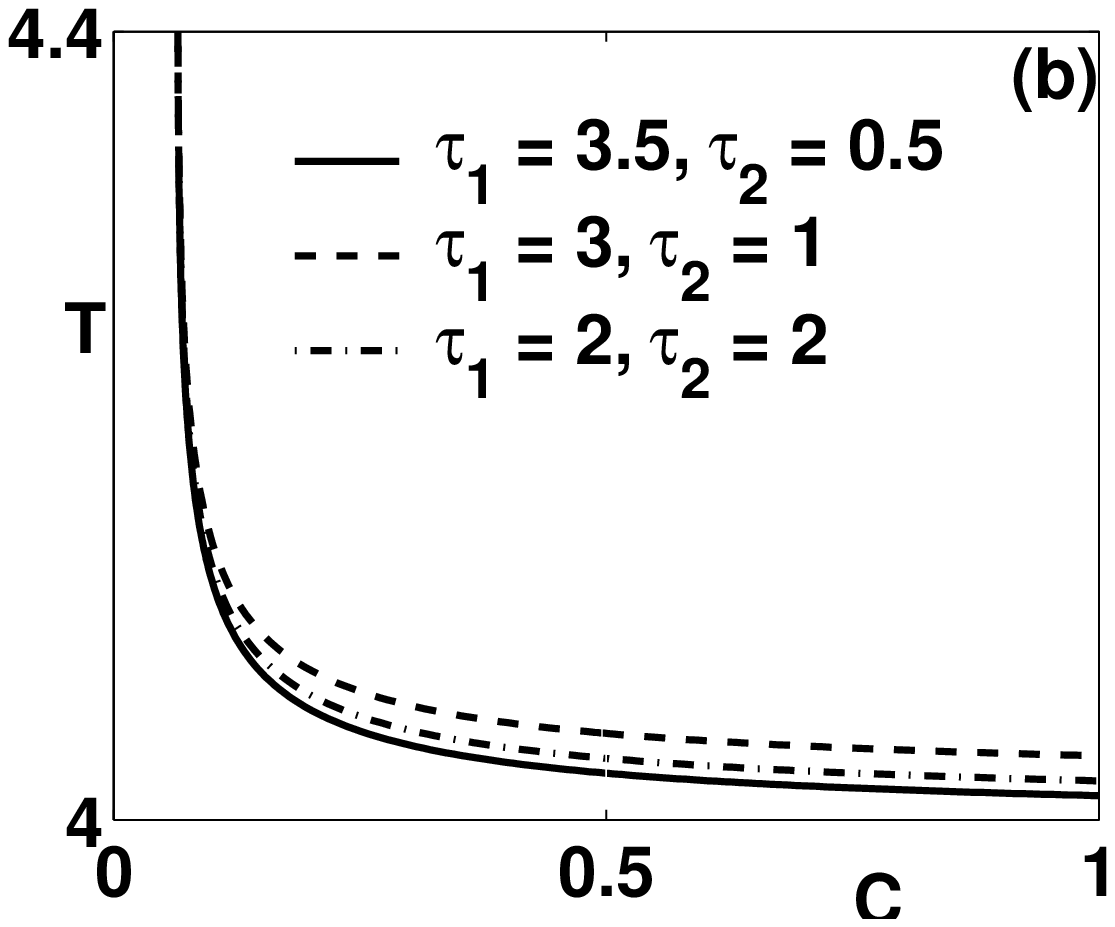}, height = 0.335\linewidth}
  \caption{Evolution of the period for the long cycle solution
    depending on the coupling 
    term. (a) Period $T$ vs. $\tau_1$ for different fixed values of
    $\tau_2$, $C = 0.5$. (b) Period $T$ vs. $C$, for different values of
    $\tau_1$ and $\tau_2$ so that $\tau_1 + \tau_2 = 4$.} 
  \label{fig:long_prd_vs}
\end{figure}

We also would like to examine the question how the
cycle period is related to the coupling
terms. The Fig.~\ref{fig:long_prd_vs}(a) represents 
several plots of the orbit period $T$ vs. $\tau_1$, while $\tau_2 =
0.5, 1, 2$ and $C = 0.5$. 

In the Fig.~\ref{fig:long_prd_vs}(b), dependence of the period on $C$ is
depicted ($\tau_2$ is the same as in (a), and $\tau_1$ is chosen so
that the sum of delays does not change). As it is expected
(cf. \cite{dah08}), $T$ increases linearly with $\tau_1$. However, it
decays with increasing $C$.

\subsection{``Short'' cycle}

\begin{figure}[H]
  \centering
  \epsfig{file = {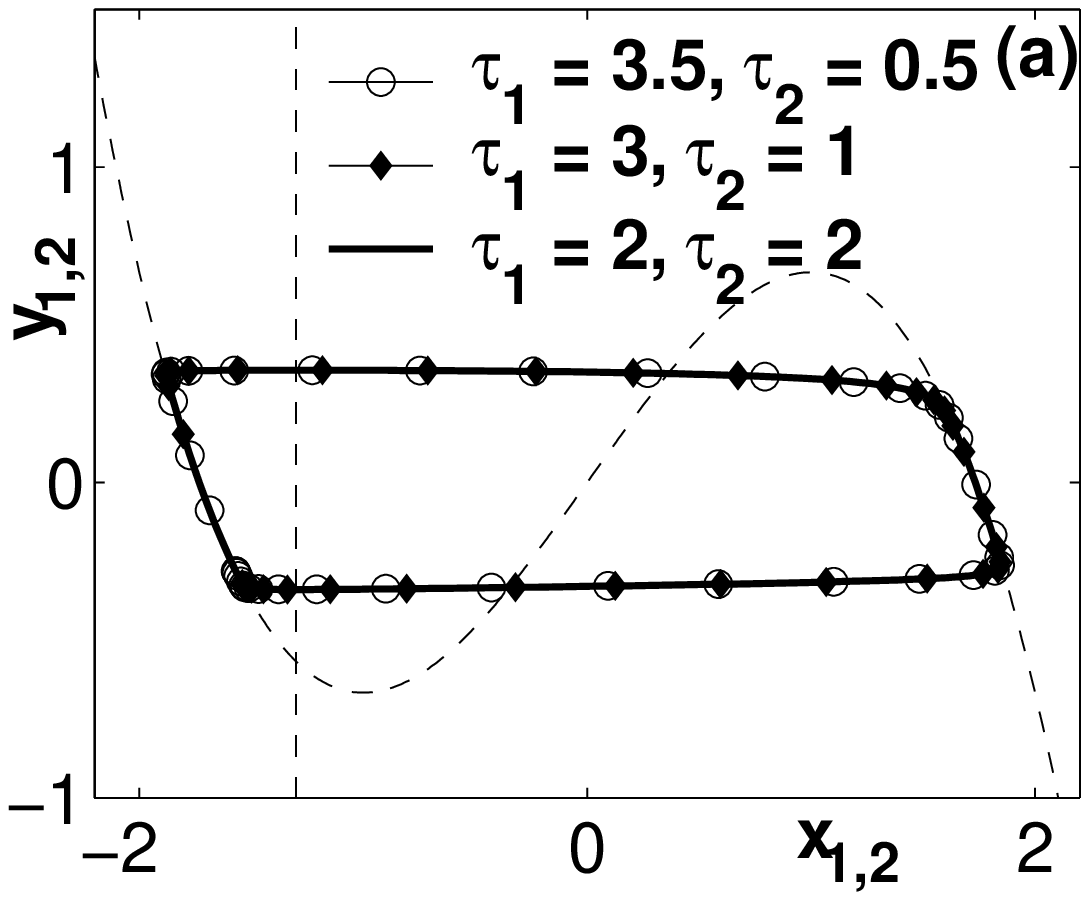}, height = 0.36\linewidth}
  \hspace{0.05\linewidth}
  \epsfig{file = {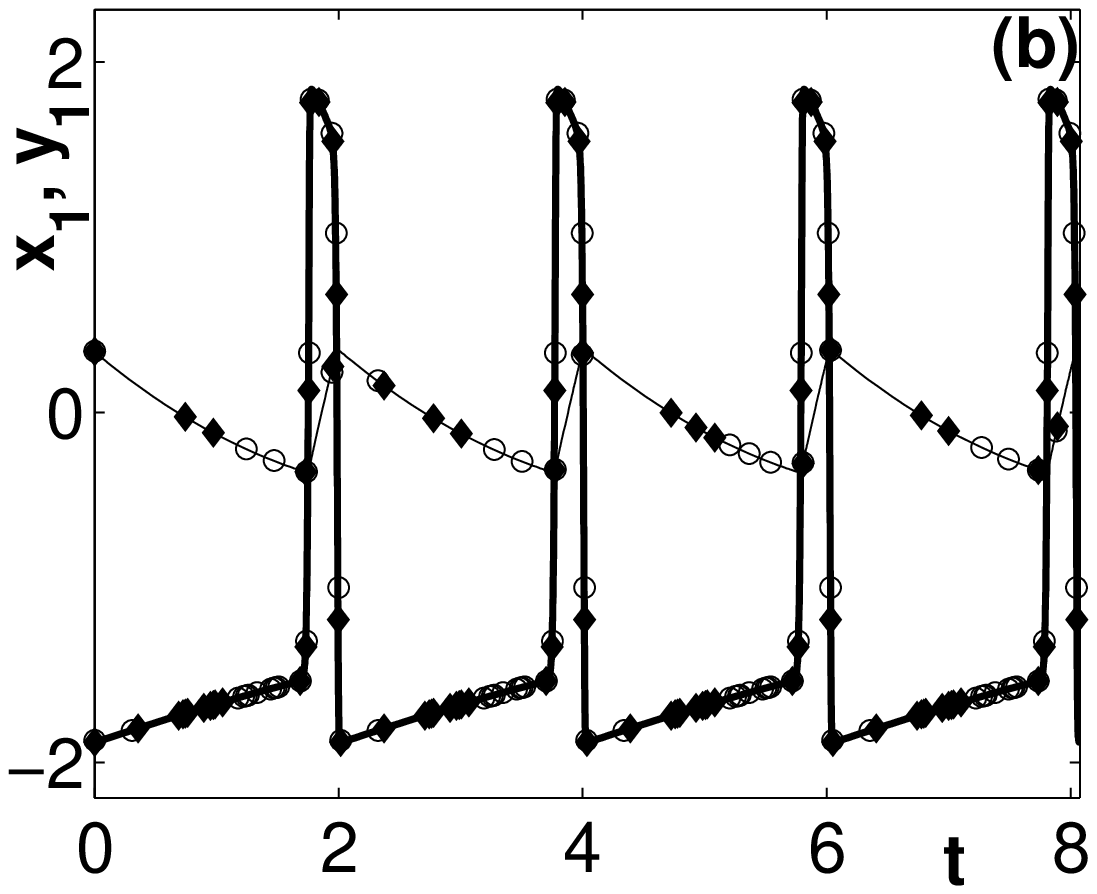}, height = 0.36\linewidth}
  \caption{Phase portrait and time series for 3 different short cycles
    with $C = 0.5$.} 
  \label{fig:short_comp}
\end{figure}

For the short cycle, the situation is almost the same. Again it is born
through a saddle-node bifurcation.
\begin{figure}[h]
  \centering
  \epsfig{file = {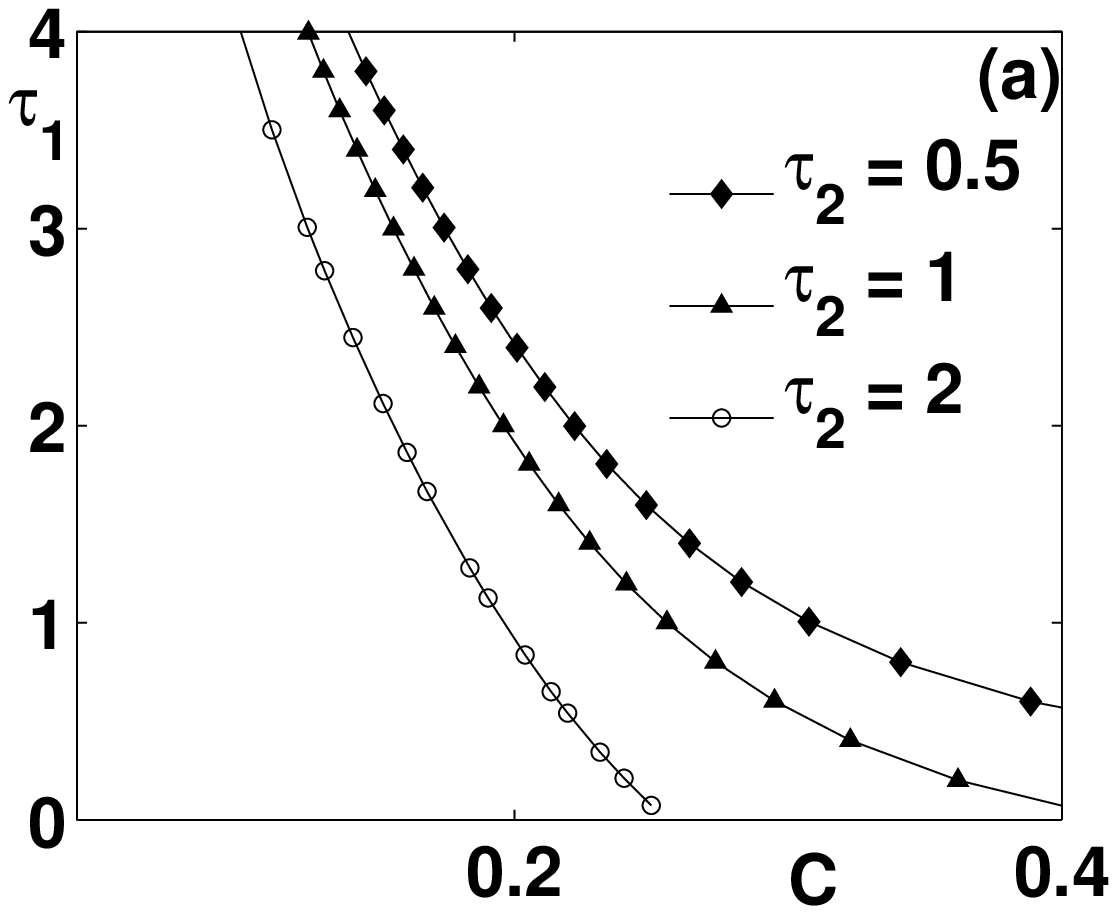}, width = 0.41\linewidth}
  \hspace{0.05\linewidth}
  \epsfig{file = {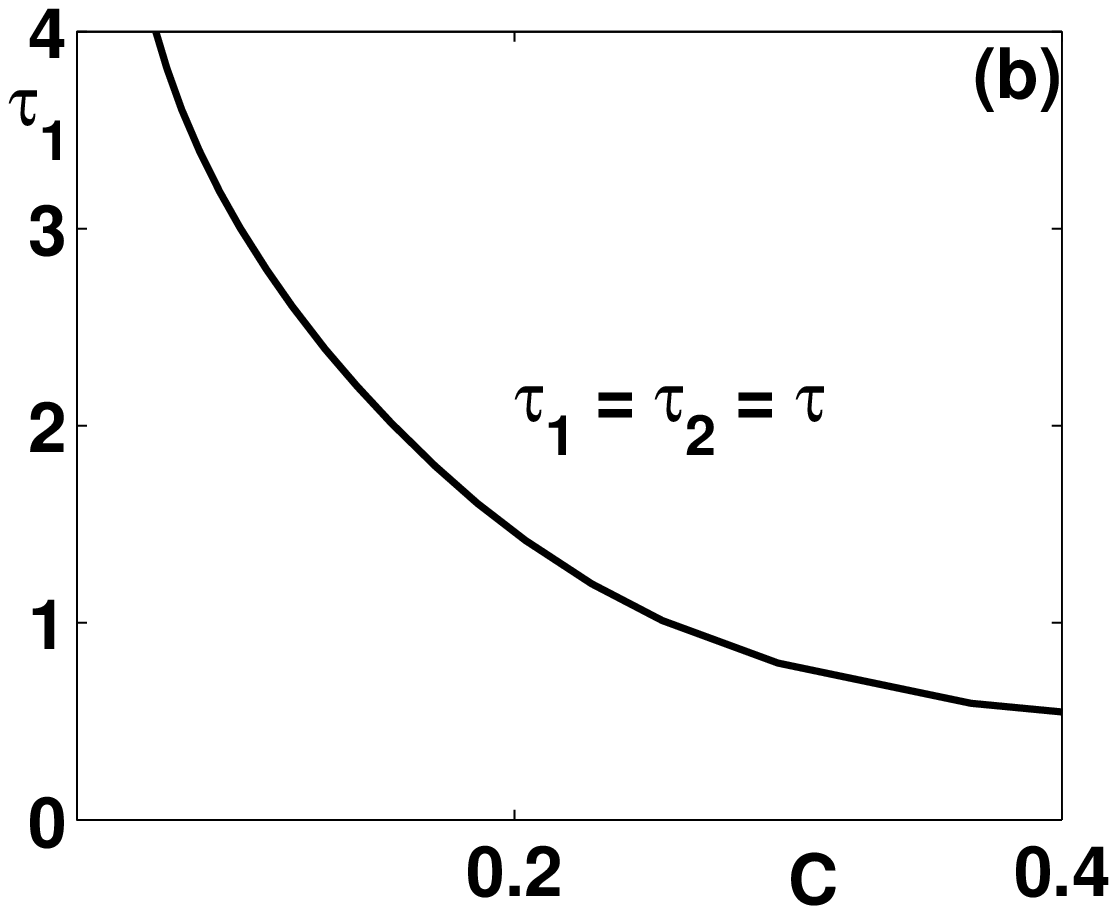}, width = 0.41\linewidth} \\
  \epsfig{file = {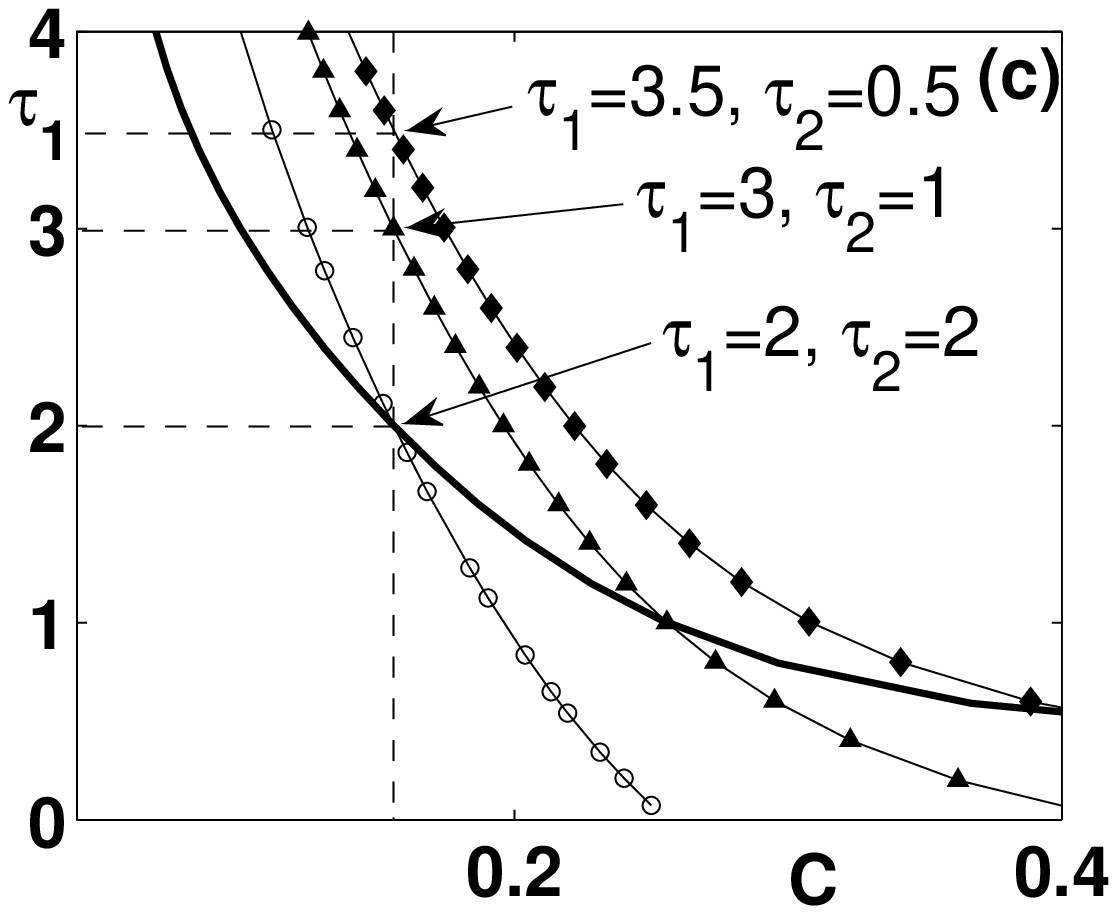}, width = 0.41\linewidth}
  \caption{Bifurcation curves for appearance of the $(\tau_1 +
    \tau_2)/2$-periodic solution (short cycle) 
    in $(C, \tau_1)$-plane. (a) Diagrams for different $\tau_2$
    values. (b) Diagram for the case of $\tau_1 = \tau_2 = \tau$. (c)
    Overlay of the figures (a) and (b).}
  \label{fig:short_bc}
\end{figure}
%
%
In the Fig.~\ref{fig:short_bc}(a), we also plot the bifurcation curves,
separating the regions of existence and absence of
the short cycle, in the $(C, 
\tau_1)$-plane (as earlier $\tau_2 = 0.5$, $\tau_2 = 1$ and $\tau_2
= 2$). Again with increasing $\tau_2$ the bifurcation curve moves to
the left, however, in comparison with the long cycle the short one
occurs for larger values of coupling strength. And after laying over
the curve for the case of equal delays $\tau_1 = \tau_2$
(Fig.~\ref{fig:short_bc}(b)), one can  
notice that the critical $C$ depends only on the delays sum (see
Fig.~\ref{fig:short_bc}(c)). The 
phase portraits and the time series for three different periodic
solutions, plotted in the Fig.~\ref{fig:short_comp}, coincide as well.

\begin{figure}[h]
  \centering
  \epsfig{file = {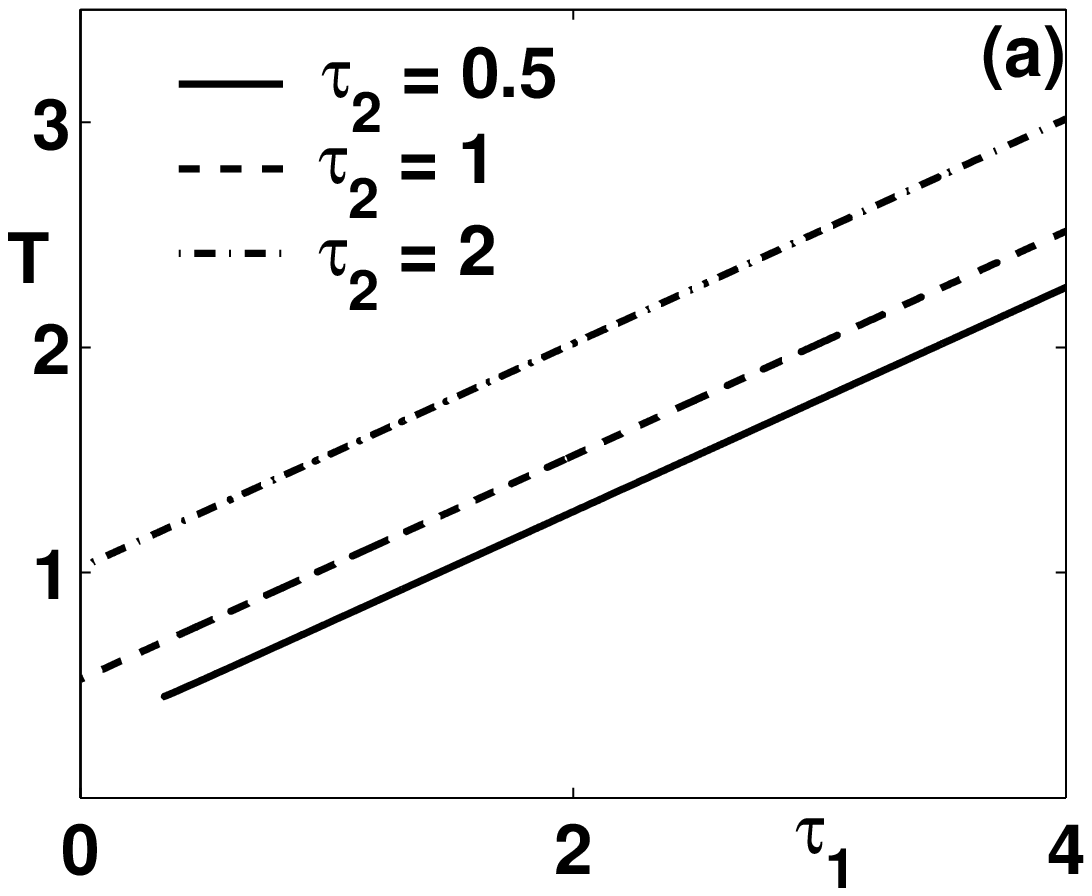}, height =
    0.335\linewidth}
  \epsfig{file = {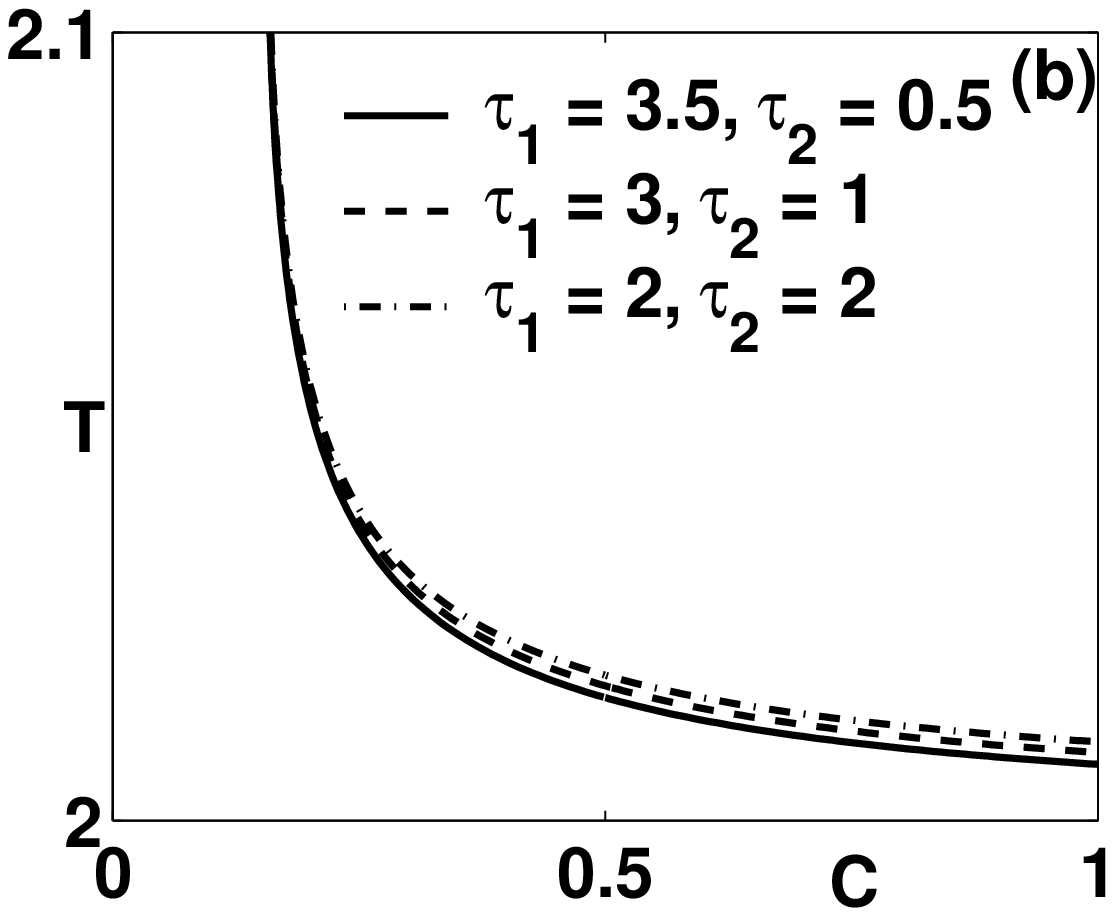}, height =
    0.335\linewidth}
  \caption{Evolution of the period for the short cycle solution
    depending on the coupling 
    term. (a) Period $T$ vs. $\tau_1$ for different fixed values of
    $\tau_2$, $C = 0.5$. (b) Period $T$ vs. $C$, for different values of
    $\tau_1$ and $\tau_2$ so that $\tau_1 + \tau_2 = 4$.}
  \label{fig:short_prd_vs}
\end{figure}

Finally, in the Fig.~\ref{fig:short_prd_vs}(a),(b), the graphs
disclosing the relation between the period and the coupling term
configuration are presented. As in the case of the long cycle, $T$ is
a linear 
function of $\tau_1$ and has a gradual decrease
on $C$.

\section{Conclusions}
In the present paper we have considered two asymmetrically
delay-coupled FitzHugh-Nagumo systems for modelling interacting
excitable neural elements. Such an ``intrusion'' gives rise
to the regular spiking in the system investigated. For sufficiently
large coupling strength and delays, one can 
observe periodic solutions of two different types (long and short
cycles),
depending on whether only one subsystem is perturbed initially or
both. The long cycle period approximately equals 
$\tau_1 + \tau_2$, while the short one has a period of about a
half of this amount. 

Furthermore, the numerical simulation, as well as the
mathematical anlysis, shows that phase portraits and time
series of these solutions do not depend on the difference of delays,
but only on their sum.

\section{Acknowledgements}
Support from DFG in the framework of Sfb 555 is acknowledged.

The authors would like to thank G.~Hiller, P.~H\"{o}vel and V.~Zykov
for fruitful discussions and remarks.



\appendix
\section{Transformation to symmetric coupling}
Consider the general system
\begin{gather}
  \label{eq:gensys1}
    \dot{x}_1 = f(x_1) + C(x_2(t - \tau_2) - x_1(t)), \\
  \label{eq:gensys2}
    \dot{x}_2 = f(x_2) + C(x_1(t - \tau_1) - x_2(t)).
\end{gather}
Without losing generality assume that $\tau_1 > \tau_2$ and denote
$\tau \equiv (\tau_1 + 
\tau_2)/2$ and $\Delta \tau \equiv (\tau_1 - \tau_2)/2$, so that
$\tau_1 = \tau + \Delta \tau$ and $\tau_2 = \tau - \Delta \tau$. Then
introducing a new function $\tilde{x}_2(t) = x_2(t + \Delta \tau)$ we
use 
$$
  x_2(t - \tau_2) = \tilde{x}_2(t - \tau)
$$ 
in
eq.~(\ref{eq:gensys1}) and rewrite the equation (\ref{eq:gensys2}) as
follows
\begin{equation*}
  \begin{split}
    \dot{\tilde{x}}_2(t) &= f(\tilde{x}_2(t)) + C(x_1(t + \Delta \tau
    - \tau_1) - \tilde{x}_2(t)) \\ 
    &= f(\tilde{x}_2(t)) + C(x_1(t - \tau) - \tilde{x}_2(t)),
  \end{split}
\end{equation*}
which leads to
\begin{equation}
  \label{eq:gensysnew}
  \begin{split}
    \dot{x}_1 &= f(x_1) + C(\tilde{x}_2(t - \tau) - x_1(t)), \\
    \dot{\tilde{x}}_2 &= f(\tilde{x}_2) + C(x_1(t - \tau) -
    \tilde{x}_2(t)).
  \end{split}
\end{equation}
This corresponds to a system with symmetric
delay coupling, and the function $\tilde{x}_2(t)$ fully coincides with
the function $x_2(t)$ of the initial problem, but with a {\em shift}
along the time axis by $\Delta \tau = (\tau_1 - \tau_2)/2$.


We also note that the inhibitor variables $y_1$, $y_2$ of the system
(\ref{eq:fhn}) depend only on $x_1(t)$ and $x_2(t)$,
respectively. Therefore, omitting them in the above analysis does not
influence the resulting conclusion.

\end{document}